\RequirePackage{fixltx2e}
\documentclass[a4paper,12pt]{article}
\usepackage{graphicx,graphics,amsmath,amssymb,rotate,textcomp,gensymb}
\usepackage{mathrsfs,float}
\usepackage[utf8]{inputenc}
\usepackage[T1]{fontenc}
\usepackage{dcolumn}
\usepackage{wrapfig}
\usepackage{caption,subcaption}
\usepackage[none]{hyphenat}
\usepackage{hyperref}
\usepackage{times}
\usepackage{multirow}
\usepackage{cite}
\usepackage{soul}

\title{Origin and structure of liquid crystalline Blue Phase III}

\author{Tanay Paul$^{1*}$, Jayashree Saha$^{1\dag}$\\
\\
\normalsize{$^{1}$Department of Physics, University of Calcutta,}\\
\normalsize{92, A. P. C. Road, Kolkata - 700009, India.}\\
\normalsize{$^*$email: tanaypaul9492@gmail.com}\\
\normalsize{Correspondence to: $^\dag$email: jsphy@caluniv.ac.in}}

\date{}

\begin{document}

\maketitle

\begin{abstract}
We report here an off-lattice NVT molecular dynamics simulation study of a system of polar chiral ellipsoidal molecules, which spontaneously exhibits Blue Phase III (BPIII), considering coarse-grained attractive-repulsive pair interaction appropriate for anisotropic liquid crystal mesogens. We have observed that suitable selection of chiral and dipolar strengths not only gives rise to thermodynamically stable BPIII but novel Smectic and Bilayered BPIII as well. Further, we have demonstrated that the occurrence of BPIII and its layered counterparts depend crucially on molecular elongation. 
\end{abstract}

\section{Introduction}
Chiral liquid crystals, in addition to broken rotational and translational (optional) symmetries, belong to the symmetry group that lacks reflection symmetry. 
Chiral liquid crystal molecules exhibit translationally disordered cholesteric phase where the nematic director precesses about the helical axis. The isotropic-cholesteric phase transition can go through a cascade of intermediate Blue Phases (BP), which have drawn widespread interest in recent technological and biological fields of research. Experimental evidences \cite{crooker, marcus, finn, koistinen} indicate that though the BPI and the BPII possess cubic structure for the double-twist cylinders, 
BPIII, the so-called `blue fog' \cite{finn, wright}, seems to be amorphous. Apparently, the BPIII does not exhibit Bragg scattering \cite{hornreich86} 
whereas it is attributed with strong optical activity along with thermodynamic stability but is lacking birefringence and having mechanical properties more like isotropic phase. 
Among chiral phases, the origin and structure of the BPIII phase still remain as a long-standing puzzle, despite extensive endeavour \cite{crooker}, because experimental explorations have been very much hindered by the occurrence of the BPIII phase in an extremely narrow temperature (or other factors affecting phase transitions) interval and optical effects are manifested only at very short wavelength range. 

There are theoretical studies on the structure of the BPIII proposing that it is a quasicrystal \cite{hornreich86, rokhsar}. Others conclude that BPIII is an amorphous phase having `spaghetti' like arrangement of double-twist cylinders (DTCs) \cite{hornreich82}, or could be a metastable state \cite{finn}. Electron micrograph experiments support an amorphous structure and thus eliminate the quasicrystal theory. A recent computer simulation study provides evidence that the structure of BPIII is basically an amorphous network of disclinations \cite{henrich}. Theoretical \cite{crooker, rokhsar, hornreich82, henrich} models proposing amorphous `spaghetti' like arrangement of double-twist cylinders and amorphous network of disclinations helped in determining phase properties. 
However, a complete understanding of these behaviours requires the information depicting molecular arrangement at the microscopic level responsible for giving rise to the BPIII, which is yet to be achieved. In the present work, we have considered systems composed of molecules interacting through chiral, dipolar and attractive-repulsive van-der-Waals' type interactions. In this Molecular Dynamics simulation study, a microscopic structure 
resembling BPIII phase 
and its layered and novel bilayered counterparts have been realized. We also have found that molecular elongation has supported more efficient self-assembly, thus has acted as a stimulating molecular feature to widen stability region. Additionally, the study of the chain-length size \cite{mtstone} of the chiral molecules is important for many technological devices \cite{azhari, varshney, stojanovic}, lipid bilayer cell-membranes \cite{skandani}, chiral drugs \cite{fu}, and physiological aspects \cite{barzilai}. 

\section{Model and Computational Details}
\begin{figure}[!b]
 \centering
 \includegraphics[width=0.65\textwidth]{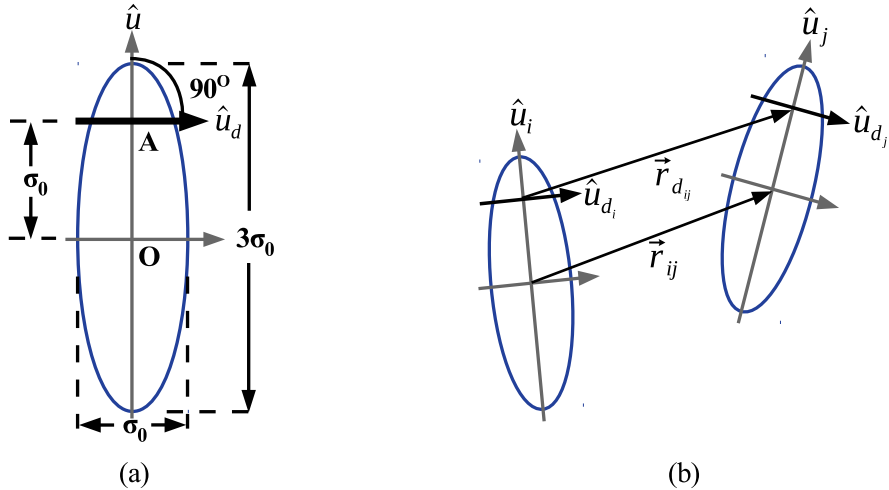}
 \caption{(a) Modeling of a polar chiral molecule of size ratio $\kappa=3.0$. A point dipole is placed at the point A, the direction of which is fixed at angle $90\degree$ with respect to the molecular long axis. (b) Relative orientations and positions of the molecules and the dipoles. 
  \label{fig:modeling}}
\end{figure}

A coarse-grained model of a polar chiral molecule having prolate ellipsoidal shape (figure: \ref{fig:modeling}a) has been used in this NVT Molecular Dynamics study \cite{allenbook, frenkelbook}. In this work, the pair potential acting between two such molecules $i$ and $j$ has been taken as \cite{memmer2k, paul19},
\begin{eqnarray}
U(\vec{r}_{ij},\hat{u}_{i},\hat{u}_{j})&=& -c\ U_{C}(\vec{r}_{ij},\hat{u}_{i},\hat{u}_{j}) + U_{dd}(\vec{r}_{d_{ij}},\hat{u}_{d_{i}},\hat{u}_{d_{j}}) \nonumber \\
& & + U_{GB}(\vec{r}_{ij},\hat{u}_{i},\hat{u}_{j}) \nonumber\\
&=& -c\ 4\epsilon(\hat{r}_{ij}, \hat{u}_{i}, \hat{u}_{j})R_{ij}^{-7}\{(\hat{u}_{i}\times\hat{u}_{j})\cdot\hat{r}_{ij}\}(\hat{u}_{i}\cdot\hat{u}_{j}) \nonumber \\
& & + \frac{1}{r^3_{d_{ij}}}[\vec{\mu}_{d_{i}}\cdot\vec{\mu}_{d_{j}}-\frac{3}{r^{2}_{d_{ij}}}(\vec{\mu}_{d_{i}}\cdot\vec{r}_{d_{ij}})(\vec{\mu}_{d_{j}}\cdot\vec{r}_{d_{ij}})] \nonumber \\
& & + 4\epsilon(\hat{r}_{ij}, \hat{u}_{i}, \hat{u}_{j})(R_{ij}^{-12}-R_{ij}^{-6}). \label{eq:1}
\end{eqnarray}
The first, second and third terms in equation \ref{eq:1} represent the chiral, dipolar and Gay-Berne \cite{gb} interactions respectively. 
Here, $\vec{r}_{ij}$ is the separation between the centers of mass of the molecules $i$ and $j$; $\hat{u}_{i}$ and $\hat{u}_{j}$ represent the orientations of the molecular long axes of respective molecules with respect to the laboratory axes (figure: \ref{fig:modeling}b). %

Here $c$ is the chirality strength parameter and the handedness of the twist depends on its sign. The form of the chiral interaction potential $U_{C}(\vec{r}_{ij},\hat{u}_{i},\hat{u}_{j})$ (first term in equation \ref{eq:1}) used in our study can directly be obtained from the multipole expansion of electrostatic interaction \cite{van-der-meer}. To incorporate the ellipsoidal shape of the molecules the orientation dependent well-depth term $\epsilon(\hat{r}_{ij}, \hat{u}_{i}, \hat{u}_{j})$ and separation term $R_{ij}=[r_{ij}-\sigma(\hat{r}_{ij},\hat{u}_{i},\hat{u}_{j})+\sigma_{0}]/\sigma_{0}$ have been taken as of Gay-Berne type \cite{gb}, where $\sigma(\hat{r}_{ij},\hat{u}_{i},\hat{u}_{j})$ is the contact distance i.e. the minimum approachable distance between two molecules. $\sigma_0$ corresponds to the contact distance at side-by-side ($\hat{u}_{i}\parallel \hat{u}_{j}$ and $\hat{u}_{i}$, $\hat{u}_{j}\perp \hat{r}_{ij}$) configuration which is also the breadth of the molecule. The term $\{(\hat{u}_{i}\times\hat{u}_{j})\cdot\hat{r}_{ij}\}(\hat{u}_{i}\cdot\hat{u}_{j})$ induces a twist 
giving rise to chiral phases.

The second term in equation \ref{eq:1} is the dipolar interaction term. The polar part of a molecule has been represented by a single terminal point dipole (figure: \ref{fig:modeling}a), positioned at $0.5\sigma_0$ distance from one terminal point of a model molecule. The orientation of the dipole has been fixed at an angle of $90\degree$ relative to the molecular long axis \cite{paul19}. 
Here, $\vec{r}_{d_{ij}}=r_{d_{ij}}\hat{r}_{d_{ij}}$ is the separation between the point dipoles fixed at two molecules $i$ and $j$ (figure: \ref{fig:modeling}b), $\hat{u}_{d_{i}}$ and $\hat{u}_{d_{j}}$ are their respective orientations relative to the simulation box; `$d$' suffix corresponds to the dipole. The dipole moment vectors of respective point dipoles are $\vec{\mu}_{d_{i}}\equiv \mu^*\hat{u}_{d_{i}}$ and $\vec{\mu}_{d_{j}}\equiv \mu^*\hat{u}_{d_{j}}$ where $\mu^*= (\mu^2/\varepsilon_{0}\sigma_{0}^3)^{1/2}$ is the magnitude of the dipole moment in reduced unit, $\varepsilon_{0}$ is the well-depth for a pair of molecules in side-by-side configuration. The long-range correction of the dipolar interaction has been taken care of by standard Reaction Field technique \cite{onsager, barker, watts}. In this technique, a sphere of a cut-off radius $r_{RF}$ is considered around a particular molecular dipole and all the dipoles outside this cut-off sphere are considered to form a dielectric continuum of dielectric constant $\epsilon_{RF}$, which produces a reaction field inside the sphere. The intensity of the reaction field applied on $i$-th molecule is given by \cite{allenbook},
\begin{equation}
 \vec{\mathscr{E}}_{i}= \frac{2(\epsilon_{RF}-1)}{2\epsilon_{RF}+1}\frac{1}{r_{RF}^3}\sum_{j\in\mathscr{R}}\vec{\mu}_j.
\end{equation}
In our study, the values of cut-off radius $r_{RF}$ and continuous dielectric constant $\epsilon_{RF}$ have been taken as $r_{RF}=0.5\times$the cubic simulation box side-length and $\epsilon_{RF}=1.5$ \cite{berardi99} respectively.

The GB parameters $\kappa$ [length to breadth ratio], $\kappa'$ [well-depth ratio $=\varepsilon_{e}/\varepsilon_{0}$; $\varepsilon_{0}$ and $\varepsilon_{e}$ being the well-depths in the side-by-side and end-to-end ($\hat{u}_{i}\parallel \hat{u}_{j}$; $\hat{u}_{i}$, $\hat{u}_{j}\parallel \hat{r}_{ij}$) configurations respectively] and the relative well depth controlling parameters $\mu$ and $\nu$ have been taken as $3.0$, $1/5$, $1$ and $2$ respectively \cite{luckhurst}. Potential cut-off radius ($r_c$) has been taken as equal to the half of the simulation box-length. To minimize computation time we have considered cut-off distances the same for both GB and chiral interactions. However, both the potential functions have been smoothed out at the cut-off boundary by shifting them by the amount $U_{\text{cut}}=U(r_c)$ \cite{allenbook}.

We have used a Leap-Frog algorithm \cite{brown} for Damped Force method (Hoover's thermostat \cite{hoover}) to solve the equations of motions in this NVT-Molecular Dynamics (MD) simulation study \cite{allenbook} which is applicable for the molecules having both translational and rotational degrees of freedom. 
Scaled density $\rho^*\left(\rho^*\equiv\frac{N\sigma_{0}^3}{V}\right)$ has been set to $0.30$ for $\kappa=3.0$ \cite{luckhurst, berardi99}. $N$ is the total number of molecules and $V$ is the simulation box volume. For each system, a well equilibrated isotropic phase has been used as the initial configuration and then the scaled temperature ($T^*\equiv k_{B}T/\varepsilon_0$, $k_B=$ Boltzmann constant) has been decreased gradually to study the phase change. For a particular system at each temperature step, the initial configuration has been a previous higher temperature equilibrium phase and simulation run of $10^6$ steps has been performed to obtain the equilibrium configuration at that temperature. To meet equilibrium criteria, at each MD step, the average energy of the system has been calculated keeping its variation with MD steps within $2\%$ Root Mean Squared fluctuation about a mean value at equilibrium. Average values have been calculated from the next $10^5$ steps after equilibration. As the twisted cylindrical domains remain oriented at random in the BPIII, so they have no biasing in twist directionality, we have used conventional cubic periodic boundary condition. In our MD simulation, the force and torque equations for all the $N$ molecules have been solved through the Leap-frog Verlet integrator method to follow the real trajectory of the system in phase space. For this reason, MD simulation is much expensive computationally \cite{mcbride-wilson, chen}. To understand the phase structure of polar chiral liquid crystal molecules, we have used an NVT algorithm in this work which is developed by us and used in our earlier works \cite{paul17, paul19}.

\vspace{-0.5cm}
\section{Results}

\begin{figure}[!b]
 \centering
 \includegraphics[width=0.85\textwidth]{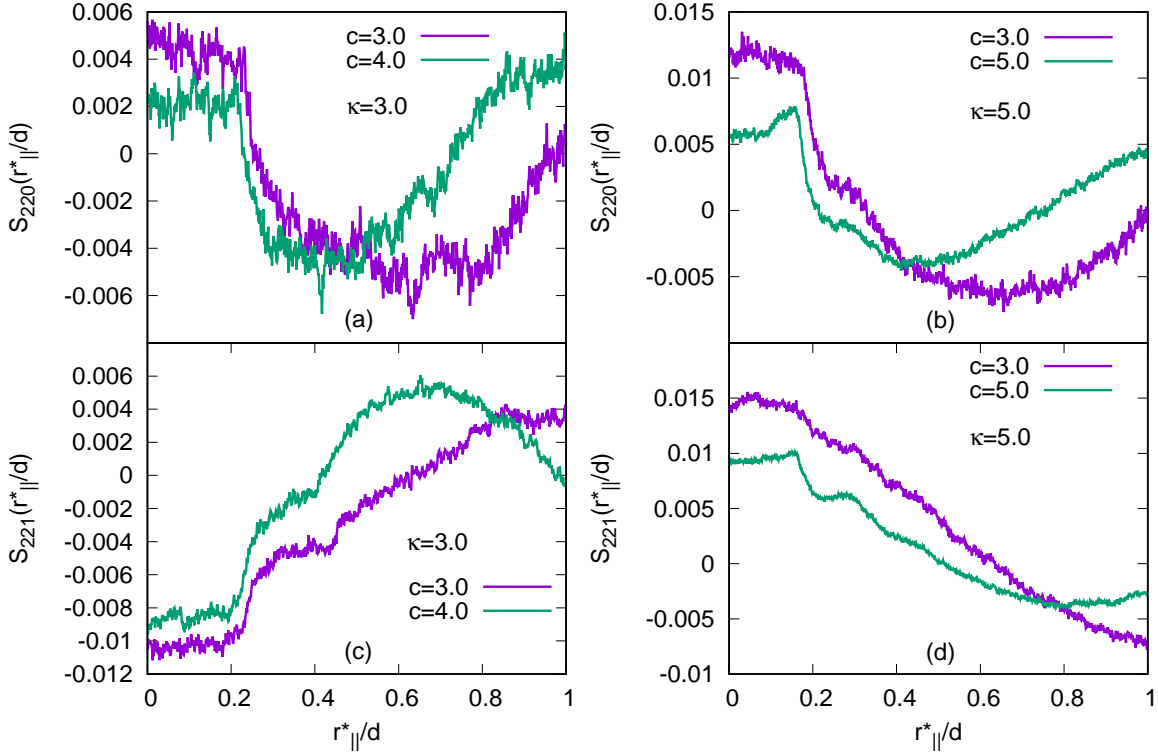}
 \caption{Plots of two longitudinal orientational correlation functions: $S_{220}(r^*_{\parallel}/d)$ for (a) $\kappa=3.0$, (b) $\kappa=5.0$ and $S_{221}(r^*_{\parallel}/d)$ for (c) $\kappa=3.0$ and (d) $\kappa=5.0$; for the systems with $N=1372$. Both the functions show Blue phase like variations with molecular separations. \label{fig:sofr}}
\end{figure}

For characterization of the phases, 
several distribution functions have been computed which have supported the formation of Blue Phase III (BPIII). 
To check system size effect we have simulated system sizes corresponding to $N=500,864,1372$ and some results for $N=2048$. The results have shown qualitatively similar phase sequence behaviour, though stabilization of a particular phase occurs at a different range of scaled temperature, usually shifted to a lower value 
for larger system size, which has been due to the finite size effects. In this paper we have presented the results of the system sizes $N=1372$ and $2048$. For all the system sizes with $\kappa=3.0$, phase properties have been studied for some selected discrete values of the chiral strength parameter $c$. 
The value of the reduced dipole moment $\mu^*$ has been set fixed at a typical value of $1.0$ \cite{paul19}. For $c=3.0$ \& $4.0$, Blue phases have been generated from a higher temperature isotropic phase (at $T^*=6.0$) by decreasing temperature. To find out the orientational correlations between the molecules, longitudinal orientational correlation functions \cite{ajstone} $S_{220}(r^*_{\parallel}/d)$ and $S_{221}(r^*_{\parallel}/d)$ have been calculated as functions of the intermolecular separation $r^*_{\parallel}$ (in units of $\sigma_0$) measured along 
a reference axis and further scaled by a 
distance $d$ which was related to the periodicity of the phase studied in Ref. \cite{memmer2k}. 
Here, these functions have been calculated considering minimum image convention and taking 
the reference axis along one of the simulation box axes and the simulation box-length as the scaling length $d$. The mathematical forms of these functions are given by,
\begin{eqnarray}
S_{220}(r^*_{\parallel}/d)&=&\frac{1}{2\sqrt{5}}\langle 3(\hat{u}_i\cdot\hat{u}_j)^{2}-1\rangle\\
S_{221}(r^*_{\parallel}/d)&=&-\sqrt{\frac{3}{10}}\langle[(\hat{u}_{i}\times\hat{u}_{j})\cdot\hat{r}_{ij}](\hat{u}_{i}\cdot\hat{u}_{j})\rangle
\end{eqnarray}
\begin{figure}[!b]
 \centering
 \includegraphics[width=0.9\textwidth]{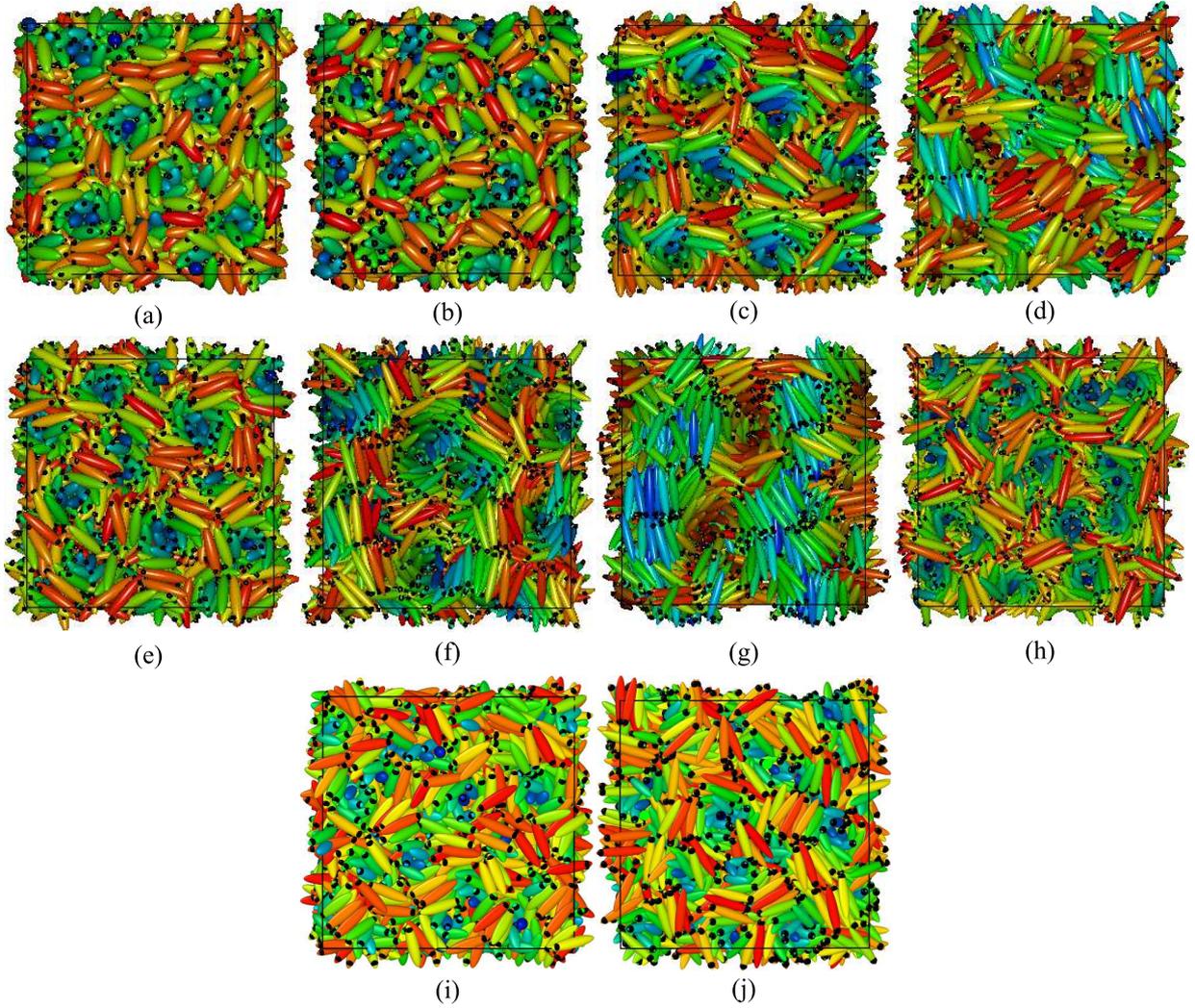}
 \caption{Different phases formed with $N=1372$: (a) BPIII with $\kappa=3.0$, $c=3.0$, $\mu^*=1.0$, (b) BPIII with $\kappa=3.0$, $c=4.0$, $\mu^*=1.0$, (c) BPIII with $\kappa=4.0$, $c=3.0$, $\mu^*=1.0$, (d) Bilayered BPIII with $\kappa=4.0$, $c=3.0$, $\mu^*=1.6$, (e) BPIII with $\kappa=4.0$, $c=4.0$, $\mu^*=1.0$, (f) BPIII, $\kappa=5.0$, $c=3.0$, $\mu^*=1.0$, (g) Bilayered BPIII, $\kappa=5.0$, $c=3.0$, $\mu^*=1.4$, (h) BPIII, $\kappa=5.0$, $c=5.0$, $\mu^*=1.0$; and with $N=2048$: (i) BPIII, $\kappa=4.0$, (j) $\kappa=5.0$, $c=5.0$, $\mu^*=1.0$. Variation of the colour refers to different molecular orientation. Dipolar positions are shown in black. 
 \label{fig:qmga}}
\end{figure}
where 
$\langle ... \rangle$ 
indicates the average over all molecular pairs separated by a distance $r^*_{\parallel}/d$ along 
the chosen reference axis. Here, the function $S_{220}$ has a maximum for two parallel molecules, whereas, the function $S_{221}$ has an extremum for two side-by-side molecules with $45\degree$ angle between their long axes and thus both help characterizing the chiral phases. The plots of $S_{220}(r^*_{\parallel}/d)$ and $S_{221}(r^*_{\parallel}/d)$ (figure \ref{fig:sofr}) show qualitatively the same variation as that of a Blue phase, i.e. depending on pitch length they vary approximately as sinusoidal functions with distance \cite{paul19, memmer2k} but the plots are not smooth, because in BPIII cylinders formed from twisted molecular arrangements are themselves twisted and are having comparatively shorter lengths, but in other Blue phases twisted molecular organization makes double-twist cylinders with straight-line symmetry axes spanning over the whole system. For other chiral phases obtained with lower $c$ values, these plots were quite smooth and showed sinusoidal type variation \cite{paul19, memmer2k} indicating the presence of definite 
orientational correlation between the molecules spreading over the whole simulation box. 
\begin{figure}[!b]
 \centering
 \includegraphics[width=0.85\textwidth]{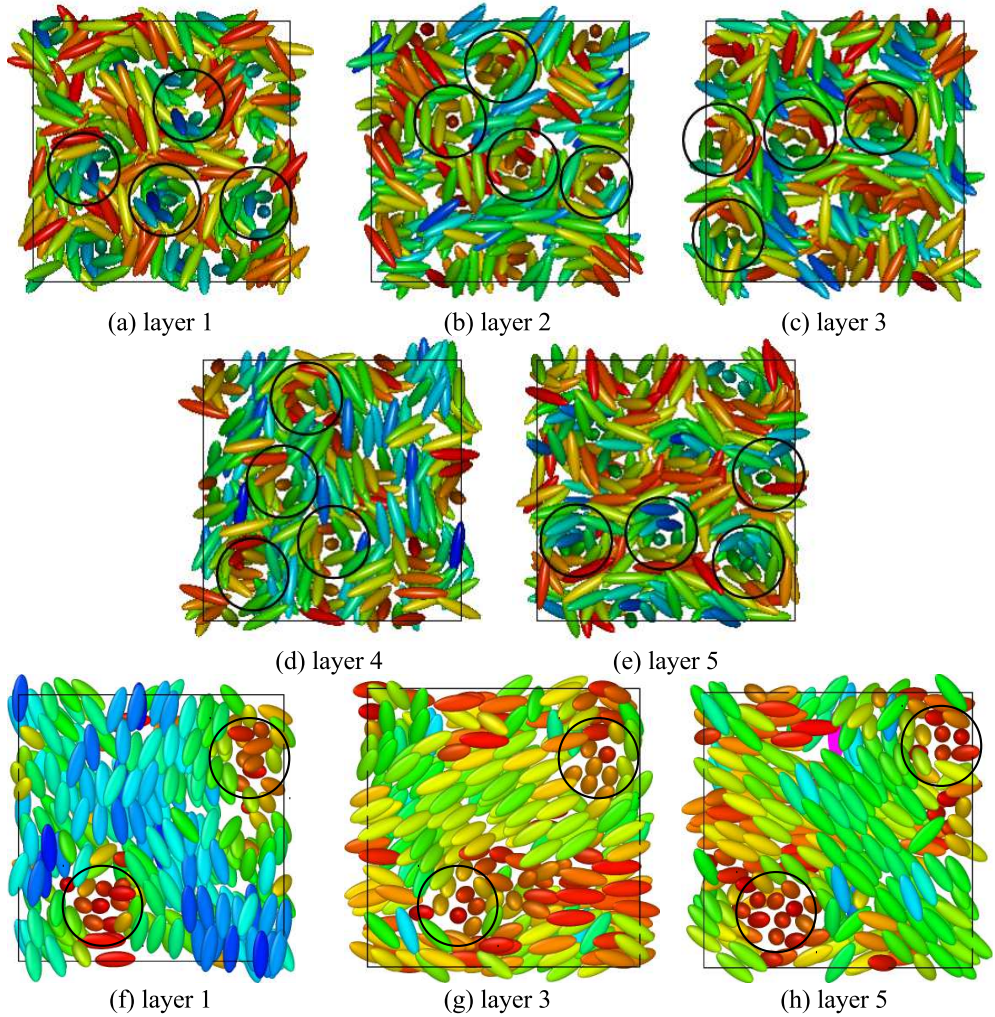}
 \caption{(a)-(e) Snapshots of the configuration in five separate consecutive layers dividing the simulation box perpendicular to $x$-axis for a system with $N=1372$, $\kappa=3.0$ obtained with $c=3.0$ at $T^*=1.8$. Circles are provided representing cross-section of some of the double-twist cylinders (DTCs) which show that double-twist cylinders are not extended straight from one box face to the opposite, unlike cubic BP phases \cite{paul19}. (f)-(h) Cross-sections of the DTCs in three layers of the BPI phase obtained for $c=1.0$, $\kappa=3.0$ and $N=1372$ (for comparison) \cite{paul19}. Dipoles are not shown for clarity. Variation of the colour of the molecules refers to different molecular orientation. 
\label{fig:xsection_layers_k3.0}}
\end{figure}
In this case, with relatively higher values of $c$, the presence of a typical variation of the orientational correlation functions with molecular separation indicates that the obtained phases are not isotropic or nematic, but a chiral one, as isotropic plots show no orientational correlation at all whereas a perfect nematic phase gives almost a straight horizontal line of high correlation value with almost no variation showing a strong parallel orientational correlation between all the molecules. 
The plots indicate these phases are not like cholesteric, BPI or BPII \cite{paul19, memmer2k}, but have a different blue phase character originated from chiral molecular organization packed within short size twisted cylindrical arrangements. Snapshots of the configurations of these phases with different $c$ and $\kappa$ values are presented in figure \ref{fig:qmga}.

Snapshot of a typical configuration obtained for $\kappa=3.0$, $c=3.0$ is shown in figure \ref{fig:qmga}a. Here in the case of BPIII these cylinders themselves are seen twisted in a random fashion. Careful observation of the configurations obtained in these cases reveals that the double-twist cylinders are not extended straight from a box face to the opposite face which occurs in case of other BPs. Snapshots depicting the configurations 
within five consecutive layers of same thickness perpendicular to one box axis are provided in figures \ref{fig:xsection_layers_k3.0}(a)-(e), where circles are drawn to show the cross-sections of some of the double-twist cylinders. In figures \ref{fig:xsection_layers_k3.0}(a)-(e) shifting of the cylindrical cross-section positions in different planes indicates cylinder twist in BPIII. Figures \ref{fig:xsection_layers_k3.0}(f)-(h) show that the encircled cross-sections of the double-twist cylinders for BPI phase, obtained with lower value of $c=1.0$ \cite{paul19}, are at same positions on each layer. Thus, in case of other BPs, i.e. for BPI or BPII, double-twist cylinders go straight from one face to the opposite face of the simulation box \cite{paul19, memmer2k}, whereas here these BPIII cylinders terminate at some intermediate plane instead of spanning over the whole box. 
\begin{figure}[!b]
 \centering
 \includegraphics[width=\textwidth]{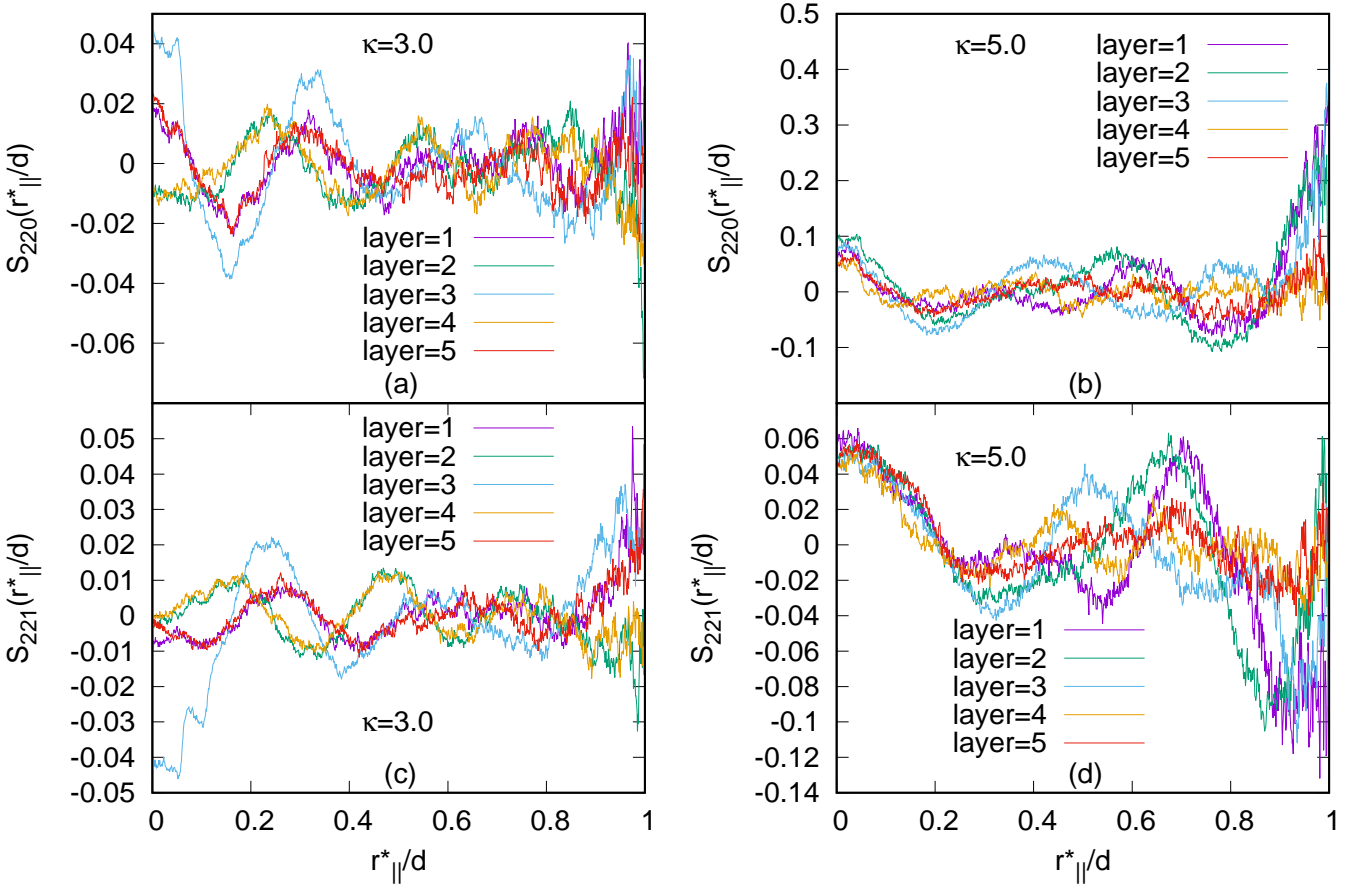}
 \caption{Plots of $S_{220}(r^*_{\parallel}/d)$ for (a) $\kappa=3.0$ and (b) $\kappa=5.0$; $S_{221}(r^*_{\parallel}/d)$ for (c) $\kappa=3.0$ and (d) $\kappa=5.0$; for the systems with $N=1372$ and $c=3.0$; dividing the simulation box in $5$ separate planar layers. $r^*_{\parallel}$ has been considered along one simulation box axis parallel to the plane. 
\label{fig:sofr_layers}}
\end{figure}

\begin{figure}[h]
 \centering
 \includegraphics[width=\textwidth]{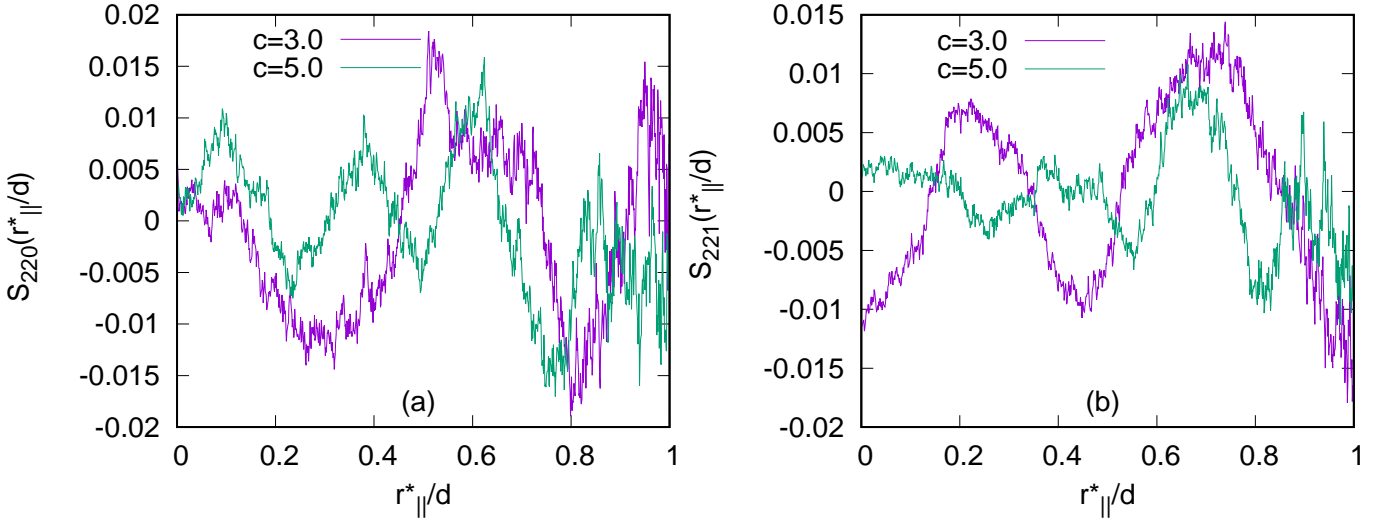}
 \caption{Plots of (a) $S_{220}(r^*_{\parallel}/d)$ and (b) $S_{221}(r^*_{\parallel}/d)$ for $\kappa=5.0$ with $c=3.0$ \& $c=5.0$ in the systems with $N = 1372$ calculated in a planar layer taking $r^*_{\parallel}$ along one simulation box axis parallel to the plane. 
\label{fig:sofr_layer2c}}
\end{figure}

To study length dependence, we have considered $\kappa$ (length to breadth ratio) $=4.0$ and $5.0$ in addition to the previous $\kappa=3.0$ case. Values of $\rho^*$ for different $\kappa$'s have been taken equal to $0.19$ for $\kappa=4.0$ and $0.12$ for $\kappa=5.0$. Again, like $\kappa=3.0$ for both $\kappa=4.0$ and $5.0$, to check chiral strength effect, we have simulated these three systems keeping $c=3.0,4.0$ and $5.0$ values respectively. 
Unlike the blue phases I and II, the double-twist cylinders here in each case are not straight, but they are tangled, as seen from the configurations. Also, at a higher value of $\kappa$, stable equilibrium blue phase has been realized at a higher value of the scaled temperature $T^*$. As for example, for systems with $N=1372$ and $c=3.0$, the typical values of $T^*$ at which equilibrium Blue phases have been obtained are $T^*=1.8$ for $\kappa=3.0$, $T^*=7.5$ for $\kappa=4.0$ and $T^*=11.0$ for $\kappa=5.0$. For further investigation, the simulation box has been divided into some planar layers perpendicular to one of the simulation box axes and then $S_{220}(r^*_{\parallel}/d)$ and $S_{221}(r^*_{\parallel}/d)$ have been calculated for all the planes separately, considering $r^*_{\parallel}$ along one of the simulation box axes parallel to the planes. Plots of these functions (figure \ref{fig:sofr_layers}) show multiple peaks, not coinciding at same points for all of the layers, but they are at different points for different layers, indicating the fact that the cross-sections of the double-twist cylinders 
in all planar layers are not at same positions. The number of peaks is more for higher values of $c$ showing lower pitch values (figure \ref{fig:sofr_layer2c}). 
This indicates that the number of double-twist cylinders increases with the increase in the chiral strength parameter $c$. These double-twist cylinders are found to be intertwined to form spaghetti-like structures \cite{hornreich82} as speculated for BPIII. 

\begin{figure}[h]
 \centering
 \includegraphics[width=0.87\textwidth]{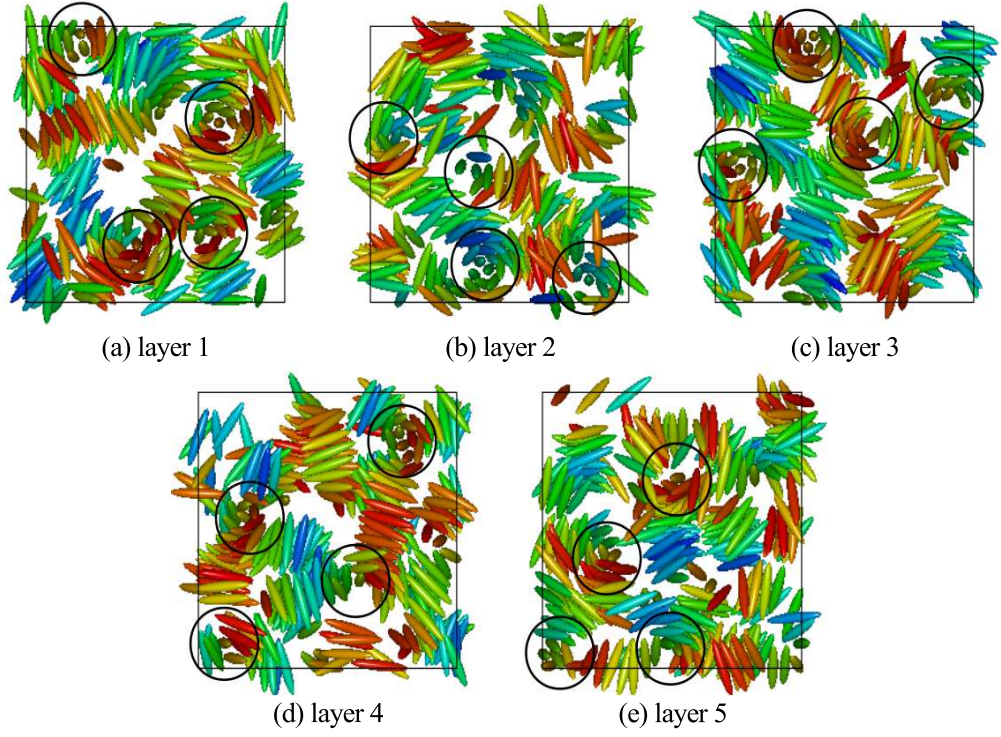}
 \caption{Snapshots of the configuration in five separate consecutive layers dividing the simulation box perpendicular to $x$-axis for a system with $N=1372$, $\kappa=5.0$ and $c=3.0$ at $T^*=9.5$. Circles are drawn representing cross-section of some of the double-twist cylinders, the non-occurrence of these circles in same positions for all layer planes indicate that double-twist cylinders are not extended straight from one box face to the opposite, unlike cubic BP phases. Dipoles are not shown for clarity. Colour variation in small smectic domains indicates twisted arrangement of molecules. 
\label{fig:xsection_layers_k5.0}}
\end{figure}

\begin{figure}[h!]
 \centering
 \includegraphics[width=0.5\textwidth]{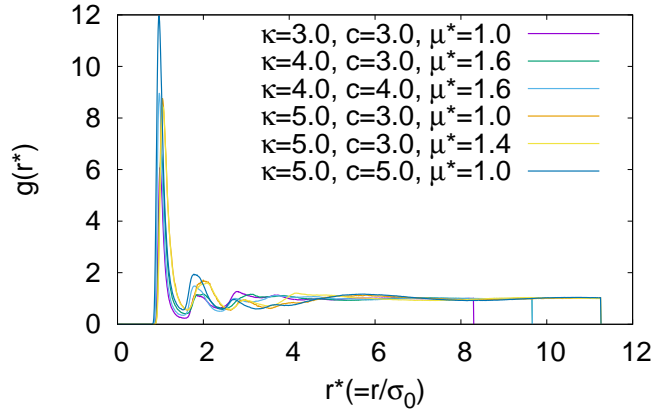}
 \caption{Plots of pair-correlation function $g(r^*)$ for different systems, where $r^*=r/\sigma_0$ is the scaled interparticle separation at a stable phase with smectic domains formed in the BPIII. 
\label{fig:gr}}
\end{figure}

With further decrease in temperature, the formation of smectic layers has started to develop in addition to the orientational arrangement of the BPIII phase. For a system with a higher value of $\kappa$, smectic layers have started forming more efficiently at a relatively higher value of the scaled temperature. As for example, for systems with $N=1372$ and $c=3.0$, the typical values of $T^*$ at which equilibrium phases with smectic domains have been obtained are $T^*=1.4$ for $\kappa=3.0$, $T^*=6.5$ for $\kappa=4.0$ and $T^*=9.5$ for $\kappa=5.0$. Smectic domains are more prominent with higher $\kappa$ values, i.e. for the systems with higher molecular lengths like $\kappa=4.0$ and $5.0$, rather than $\kappa=3.0$. Snapshots of the configuration obtained with $\kappa=5.0$ are presented in figure \ref{fig:xsection_layers_k5.0} dividing the simulation box into five planar layers perpendicular to $x$-axis for visualization of the twist of the double-twist cylinders across the planar layers and formation of layered domains. The plots of the pair distribution function $g(r^*)=V\langle \sum_i \sum_{j\ne i} \delta (r^*-r^*_{ij})\rangle /N^2$ ($r^*$ is the scaled intermolecular separation $=r/\sigma_0$) for these phases (figure: \ref{fig:gr}), at respective scaled temperature $T^*$ at which these smectic phases are stable, show short range positional order without any long range positional order.


\begin{figure}[h]
 \centering
 \includegraphics[width=\textwidth]{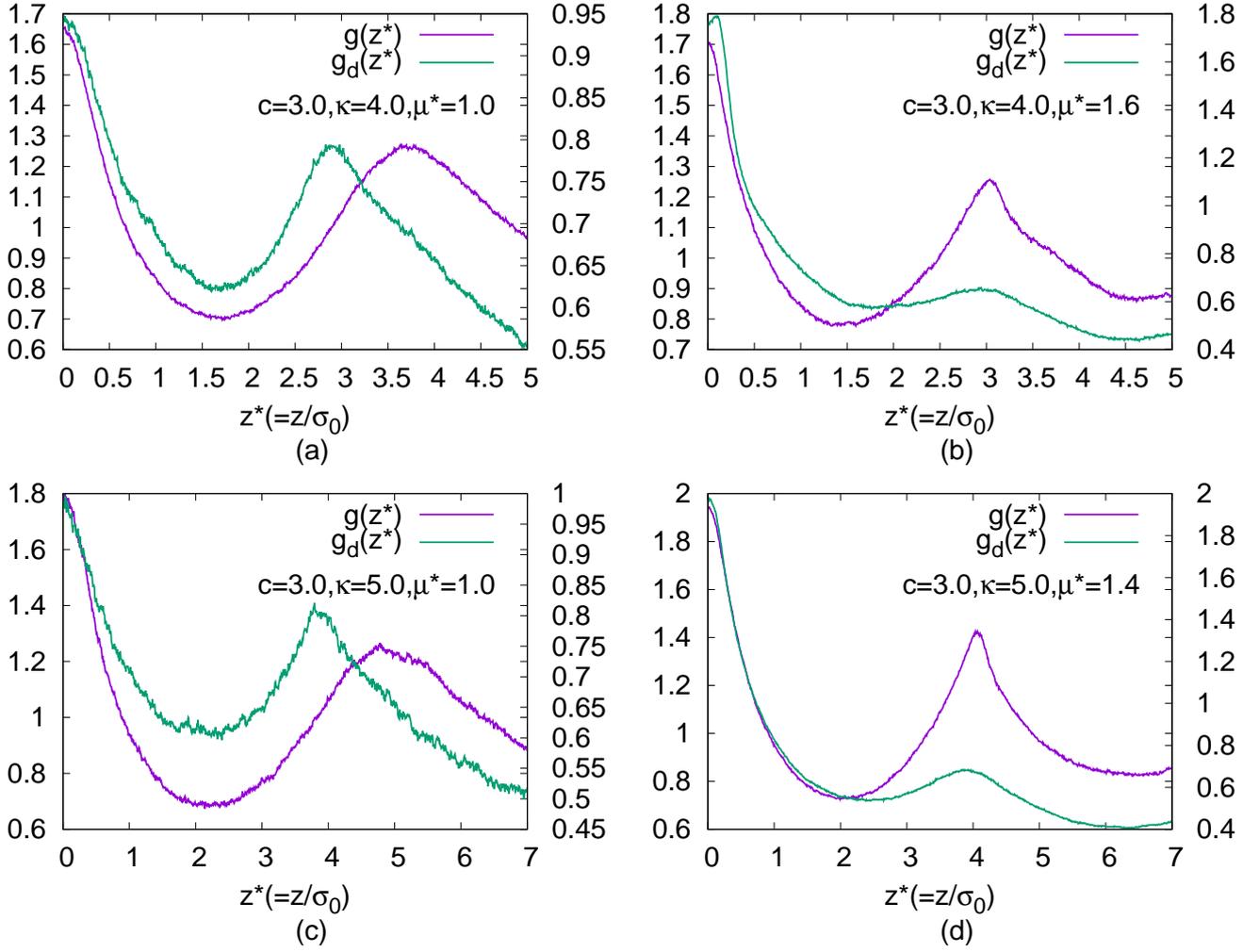}
 \caption{Plots of $g(z^*)$ and $g_d(z^*)$ calculated in the systems with $N=1372$ in small cylindrical domains around each molecule: for (a) $\kappa=4.0$, $c=3.0$, $\mu^*=1.0$; (b) $\kappa=4.0$, $c=3.0$, $\mu^*=1.6$; (c) $\kappa=5.0$, $c=3.0$, $\mu^*=1.0$; (d) $\kappa=5.0$, $c=3.0$, $\mu^*=1.4$. For relatively higher value of the dipole moment $\mu^*$ with same $\kappa$ and $c$, the peaks of comparable heights for both the functions arise in alternate positions indicating the presence of bilayered domains inside BPIII. 
\label{fig:gofzdom}}
\end{figure}

\begin{figure}[h!]
 \centering
 \includegraphics[width=0.4\textwidth]{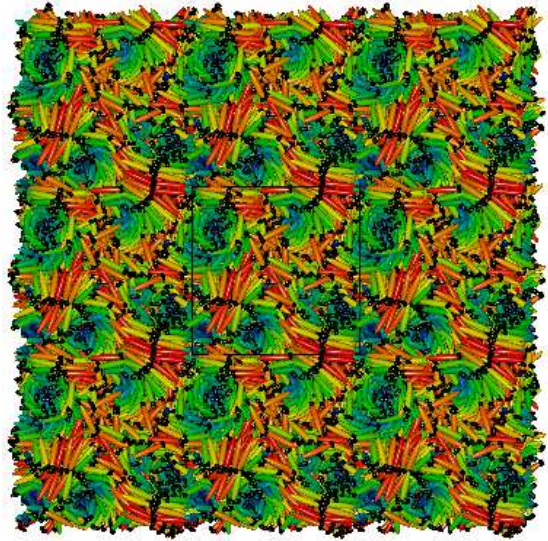}
 \caption{Visualization of Bilayered BPIII obtained in a system with $N=1372$, $c=3.0$, $\kappa=5.0$, $\mu^* = 1.4$ at $T^*=9.0$ and repeated by identical images as given by periodic boundary conditions. Molecules are shown in ellipsoids having axis ratio of $1:1:5$ and dipoles by small black dots. View of the cross-section perpendicular to one box-face is shown. 
\label{fig:qmgarpt_biBPIII}}
\end{figure}

Additionally, we have checked the effect of dipole strength on the layer formation \cite{paul19, paul17}. The presence of dipoles gives more stability to the smectic layers. Interestingly, for a higher value of $\mu^*$ novel bilayered smectic BPIII (figure \ref{fig:qmga}d and \ref{fig:qmga}g) has formed, by decreasing temperature from a relatively higher temperature phase. For a system with $N=1372$ and $c=3.0$, by decreasing the scaled temperature from $T^*=6.5$ to $T^*=6.0$ bilayered arrangement in the smectic domains has formed where $\kappa=4.0$ and $\mu^*=1.6$ (fig: \ref{fig:qmga}d), whereas in a system with $\kappa=5.0$ and $\mu^*=1.4$ (fig: \ref{fig:qmga}g) decreasing the temperature from $T^*=9.5$ to $T^*=9.0$ bilayered arrangement has formed in the smectic domains. For $\kappa=3.0$ formation of the bilayer is not so prominent. Plots of $g(z^*)$ and $g_d(z^*)$ (where the former is the pair correlation function for the molecular centers of mass as a function of the projection ($z^*=z/\sigma_0$) of separation vector along the axis of a cylindrical domain considered around each molecule and the latter is the same for dipolar positions) have been drawn (figure \ref{fig:gofzdom}) by calculating in the same fashion as $g(r^*)$ plot. Thus, $g(z^*)=\langle \delta (z^*-z^*_{ij})\rangle /\pi R^2 \rho^*$ (where $R$ is the radius of the cylindrical sampling region) and $g_d(z^*)$ has been calculated in the same way for the dipolar positions considering cylindrical domain of the same size. The peaks of the function $g(z^*)$ indicate the presence of the molecular centers of mass in planar layers perpendicular to the symmetry axes of the cylindrical sampling regions and similarly, the peaks of the function $g_d(z^*)$ indicate the presence of the dipolar positions in similar planar layers. For $\kappa=4.0$ (figure \ref{fig:gofzdom}a and \ref{fig:gofzdom}b) cylindrical domain of length $5\sigma_0$ and radius $4\sigma_0$ has been taken. Those values for $\kappa=5.0$ (figure \ref{fig:gofzdom}c and \ref{fig:gofzdom}d) are $7\sigma_0$ and $5\sigma_0$ respectively. For both the $\kappa$ values, the peaks of comparable heights for both the functions occur at nearly same positions for lower values of $\mu^*$ (figure \ref{fig:gofzdom}a and \ref{fig:gofzdom}c) indicating non-bilayer arrangements, but when the value of $\mu^*$ increases the peaks of comparable heights for $g_d(z^*)$ occur alternately with the peaks of $g(z^*)$ (figure \ref{fig:gofzdom}b and \ref{fig:gofzdom}d) indicating the presence of small bilayered smectic domains in the BPIII. 
Due to the smaller size of the smectic domains, we have obtained fewer peaks in these plots. 
A snapshot of the bilayered BPIII is presented in figure \ref{fig:qmgarpt_biBPIII} by periodic repetition of the simulation box, where bilayered domains can be identified with the dipolar ends of the molecules (shown in black dots) of two adjacent smectic layers clustered together.

\section{Conclusion}

Chirality, which is a very important phenomenon occurring in various fields of nature, when added to liquid crystal systems generates many phases featuring fantastic and remarkable properties and having wide applicability in technology, medical science, agriculture and chemical industry. Blue phases are one of this kind. While other liquid crystalline phases have rotational and translational symmetries, chiral Blue phases (BP), the so-called `crystalline liquids', have additional symmetries of conventional solid crystals i.e. cubic crystal symmetries. Being a member of liquid crystal family, this property of BP makes them interesting both as a curious subject for scientific studies and a material of immense potential for industrial use. Theoretical and computer simulation studies on BPI and BPII have proposed that for relatively high chiral strength, the simple helical structure of the cholesteric phase is energetically less stable locally than a `double-twist cylinder' structure. In a double-twist cylinder, local directors rotate simultaneously about any radius of the cylinder and the local director is parallel to the cylinder symmetry axis at its center. Such double-twist cylinders do not fit in three-dimension to fill the whole space and thus disclinations or defects are formed. Among three types of BP observed experimentally without electric field, in BPI, the arrangement of these disclinations is body center cubic and in BPII is simple cubic. However, BPIII, the so-called `blue fog' \cite{wright}, is amorphous, the structure of which is not very clear till date. Our simulation work supports the theoretical proposition of the `spaghetti' like arrangement of the double-twist cylinders as a model of BPIII and at the same time provide microscopic description of the molecular arrangement. 
In this simulation study, we have found that higher chiral strength in the system induces inter-twinning of double-twist cylinders which eventually gives rise to a structure resembling BPIII phase. Additionally, we have found that higher values of dipole moment can induce novel bilayer arrangement in the smectic BPIII. Change in phase properties with increase in molecular length has also been studied. Our observation is that the Isotropic-BPIII transition temperature is higher for a system with higher molecular length. Greater the molecular length, the required value of $c$ is lower at which BPIII is formed. We also have observed that molecular elongation favours the efficacious formation of the smectic BPIII, whereas, increased dipolar strength is the key to give rise to novel bilayered BPIII. 
The focus of our work is to find out the proper contribution of various physical microscopic interactions responsible for the realization of BPIII and its novel layered counterparts. 
Consequently, this work indicates a way that may help to minimize present experimental hurdle due to significantly narrow stability region \cite{crooker} of the BPIII phase. We hope, the present coarse-grained simulation study will help in the fundamental qualitative understanding of the molecular level arrangement in the BPIII and the coveted structure-property relationship. 

\section{Acknowledgement}

T.P. gratefully acknowledges the support of Council of Scientific \& Industrial Research (CSIR), India, for providing Senior Research Fellowship. This work is partly supported by the UGC-UPE scheme of the University of Calcutta.


\textbf{\large Author contributions statement}

Jayashree Saha supervised the work. Both the authors, Tanay Paul and Jayashree Saha, contributed to the design of the work and analysis of the data and wrote the paper. Tanay Paul carried out the simulations.

\textbf{\large Competing interests statement}

The authors declare no competing interests.
\end{document}